\documentclass[journal]{IEEEtran}

\usepackage{mathptmx,mathtools,mathrsfs}
\usepackage{amsmath,amssymb,amsfonts,amsthm}

\usepackage{graphicx,caption,subcaption}
\usepackage{textcomp,multirow}
\usepackage{hyperref}
\usepackage{booktabs}
\usepackage{xcolor}
\usepackage{colortbl}

\usepackage{algpseudocode}
\usepackage{algorithm}
\usepackage{cite}

\newcommand{\gl}{g_{\rm{MMPS}}}
\newcommand{\Rl}{\mathcal{R}_{\rm{MMPS}}}
\newcommand{\gq}{g_{\rm{ELLP}}}
\newcommand{\Rq}{\mathcal{R}_{\rm{ELLP}}}
\newcommand{\Rel}{\mathcal{R}_e}

\theoremstyle{definition}

\newtheorem{rem}{Remark}

\begin{document}

\title{Efficient MPC for Emergency Evasive Maneuvers, Part I: Hybridization of the Nonlinear Problem}

\author{Leila~Gharavi,
        Bart~De~Schutter,~\IEEEmembership{Fellow, IEEE}
        and~Simone~Baldi,~\IEEEmembership{Senior Member, IEEE}
\thanks{Leila~Gharavi and Bart~De~Schutter are with the Delft Center for Systems and Control, Delft University of Technology, 2628 CD Delft, The Netherlands (e-mails: \texttt{L.Gharavi@tudelft.nl}; \texttt{B.DeSchutter@tudelft.nl}).}
\thanks{Simone~Baldi is with the School of Mathematics, Southeast University, Nanjing 21118, China (e-mail: \texttt{103009004@seu.edu.cn}).}}

\markboth{Sumbitted to Journal,~Vol.~00, No.~0, October~2024}%
{Shell \MakeLowercase{\textit{et al.}}: Bare Demo of IEEEtran.cls for IEEE Journals}

\maketitle

\begin{abstract}
Despite the extensive application of nonlinear Model Predictive Control (MPC) in automated driving, balancing its computational efficiency with its control performance and constraint satisfaction remains a challenge in emergency scenarios: in such situations, sub-optimal but computationally rapid responses are more valuable than optimal responses obtained after long computations. This paper introduces a hybridization approach for efficient approximation of the nonlinear vehicle dynamics and of its non-convex constraints, e.g., arising during emergency evasive maneuvers. Hybridization, i.e.~, the use of hybrid systems modeling, allows to reformulate the nonlinear MPC problem as a hybrid MPC problem. Max-Min-Plus-Scaling (MMPS) hybrid modeling is used to approximate the nonlinear vehicle dynamics. Meanwhile, different formulations for constraint approximation are presented, and various grid-generation methods are compared to solve these approximation problems. Among these, two novel grid types are introduced to structurally include the influence of the nonlinear vehicle dynamics on the grid point distributions in the state domain. Overall, the work presents and compares three hybrid models and four hybrid constraints for efficient MPC synthesis and offers guidelines for implementation of the presented hybridization framework in other applications.
\end{abstract}

\begin{IEEEkeywords}
Hybridization framework, Model predictive control, Evasive maneuvers, Vehicle control
\end{IEEEkeywords}

\IEEEpeerreviewmaketitle

\section{Introduction} 
\IEEEPARstart{M}{odel} predictive control (MPC) has become increasingly popular in automated driving research over the past few decades~\cite{Stano2022}. This is mainly due to its capability to handle constraints and its ability to adapt to the system by performing controller synthesis in a rolling-horizon optimization-based manner. However, high computation loads remain a major obstacle towards real-time implementation of MPC for higher levels of automation. In particular, Level 4 and Level 5 of automation defined by the Society of Automated Engineers (SAE)~\cite{SAE} must be able to handle hazardous scenarios without any intervention from the human driver. Clearly, in such critical situations, sub-optimal but computationally rapid responses are more valuable than optimal responses obtained after long computations. Thus, improving the computational efficiency of MPC in critical scenarios remains a crucial challenge.

Several lines of research have been investigated to deal with this challenge: suggested approaches to increase computational efficiency include decoupling the lateral and longitudinal vehicle dynamics~\cite{Huang2020} or using ad-hoc kinematics and dynamics~\cite{Chowdhri2021}. Partly-related research lines have looked at how model fidelity affects the control performance during critical maneuvers in limits of friction~\cite{Subosits2021} or around drift equilibria~\cite{Patterson2021}.

Another line of research has been studying computationally more efficient solutions to the nonlinear optimization problem e.g., via new numerical algorithms~\cite{Guo2020} or offline explicit solutions~\cite{Metzler2021}. Nevertheless, Tavernini~et~al.~\cite{Tavernini2019} demonstrated that the offline explicit MPC approach does not yield significant computational improvements. Adaptive weights, adaptive prediction horizon~\cite{Oh2022} or adaptive sampling times~\cite{Brudigam2023} have also been examined, which can sometimes improve computational efficiency although Wurts~et~al.~\cite{Wurts2022} argue that varying sampling times can increase the computational burden due to the resulting change in integration points in the prediction horizon.

Switching-based control designs are another line of research for computational efficiency of MPC, for instance, by switching among different prediction models~\cite{Rokonuzzaman2022}. Nevertheless, there is often no systematic way to define  a good switching strategy, as the switching can be defined in different ways such as switching to a higher-fidelity model in case of uncontrollable error divergence~\cite{Zhang2015}, or switching among different drifting/driving modes~\cite{Zhao2021}. In this sense, a more systematic framework that covers switching-based design as a special case is hybridization~\cite{Asarin2007}. Hybridization refer to approximating the control optimization problem using a hybrid systems formulation incorporating both continuous and discrete dynamics~\cite{Lunze2009}. Hybridization is equivalent to breaking down a nonlinear possibly complex form into multiple modes with lower complexity, each mode being valid in a local activation region. By this approach, nonlinearity is traded with the introduction of discrete dynamics, representing the switching among the different modes of the system~\cite{Heemels2001}. 

Hybridization has been used to improve the computational speed in various applications~\cite{Suzuki2013,Khanmirza2016,Sun2020}. In the automated driving literature, different approaches to hybridize the vehicle dynamics include representing the nonlinear tire forces by a piecewise-affine function~\cite{Dicairano2013,Guo2020,Jagga2018}, using a grid-based linear-parameter-varying approximation~\cite{Corno2021}, or using a hybrid equivalent state machine~\cite{Amir2017}. Nevertheless, to the best of our knowledge, hybridization has not yet been incorporated into emergency evasive maneuvers and/or highly-nonlinear vehicle dynamics. For example, the hybridization in~\cite{Sun2019} via a Mixed-Logical-Dynamical (MLD) formalism~\cite{Bemporad1999} is only valid at low-speeds where vehicle nonlinearity can be neglected and the coupling between lateral and longitudinal vehicle dynamics is weaker.

Indeed, in addition to the nonlinear vehicle model that enters the MPC problem as equality constraint, another crucial source of nonlinearity in the control optimization problem is caused by the physics-based inequality constraints such as handling and tire force limits that are generally non-convex. The hybridization problem in MPC must necessarily involve both model and constraint approximations, which is often neglected in the literature. Despite some similarities, there are clear distinctions in the two resulting hybridization problems that must be taken into account. 

Among different hybrid modeling frameworks, Max-Min-Plus-Scaling (MMPS) systems~\cite{DeSchutter2020} do not require to explicitly represent the activation regions, which simplifies the approximation by significant reduction of the number of decision variables. For this reason, the MMPS approach is the one adopted in this work. As its name suggests, MMPS formulation represents a function using only (and possibly nested) max, min, adding and scaling operators. Kripfganz~\cite{Kripfganz1987} showed that any MMPS function can also be equivalently represented by the difference of two convex MMPS functions, which can increase computational tractability.

Physics-based non-convex constraints have been dealt with in different ways. For instance, \cite{Perez2012} considers the convex hull of the non-convex polyhedral constraints and disregards non-optimal solutions using the binary search tree of~\cite{Tondel2003}. In reachability analysis, \cite{Kochdumper2020} computes an inner-approximation of the feasible region using an outer approximation of the reachable sets. 

Lossless or successive convexification is a common approach to deal with non-convex constraints, as often considered in real-time trajectory planning~\cite{Miao2021,Scheffe2021,Acikmese2013}. However, the real-time capability of the convexification method is a crucial and non-trivial aspect, since the non-convex constraints imposed by the environment are changing in each control time step.

Convexification problem can be solved offline only when the constraints are known to be fixed. In some applications such as path planning in cluttered environments, it is important to find a feasible region for the next control time step, which translates into finding the largest convex subset of a given cluttered feasible region~\cite{Deits2015}. Nevertheless, a generic offline convexification problem can be obtained by approximating a non-convex region by a union of convex subregions. As defining these subregions manually is unpractical~\cite{Okamoto2019}, approaches from computational geometry have been proposed. For instance, it has been shown that convexification is analogous to the NP-hard problem of Approximate Convex Decomposition~\cite{Yao1996} with applications to shape analysis~\cite{Xu1996} or decision region in pattern recognition~\cite{Yao2014}. Indeed, the recent advances in this field have been tailored more and more toward the specific needs of pattern recognition. For example, more emphasis is given on shape analysis by concavity matrices~\cite{Wei2022}: however, in critical automated driving scenarios, it is rather important to analyze the approximations inaccuracy with respect to the distance to the non-convex boundary. Existing methods in this sense are mainly tailored for non-convex polyhedral regions~\cite{Bulbul2009}, but several physics-based constraints arising during critical maneuvers are not polyhedral.

In practice, hybridization has rarely been considered for highly complex vehicle models; e.g.,\ to the best of our knowledge, there are no studies that include hybridization of the coupled longitudinal and lateral vehicle dynamics. Moreover, controlling evasive maneuvers in critical scenarios requires a systematic analysis of the vehicle model complexity and the resulting computation trade-off, which has not been conducted as far as we are aware. 

In this paper, we provide a comparison benchmark to analyze and improve the computational performance of MPC optimization problem for vehicle control in critical high-velocity scenarios using hybrid formulation of the control optimization problem. This benchmark is divided in two parts: the first part is dedicated to the hybridization of the MPC via approximating the constraints, i.e., prediction model and physics-based constraints, whereas the second part investigates the improvements of the resulting hybrid MPC controller in comparison with the original nonlinear MPC controller. 

The current paper contributes to the state-of-the-art by:
\begin{itemize}
	\item presenting of a novel hybrid approximation of the system using an MMPS formulation,
	\item developing a new generalized formalism for constraint approximation problem including an approach based on a polytopic definition of the regions by an MMPS function, and comparing the resulting approximations with two methods from the literature,
	\item introducing two trajectory-based grid generation method for model approximation,
	\item investigating grid-based numerical solutions of the model and constraint approximation with respect to the grid behavior, and
	\item presenting a novel benchmark for evaluating and comparing the computational efficiency of various nonlinear MPC controllers.
\end{itemize}

The paper is organized as follows: Section~\ref{sec:back} covers the preliminary definitions of the model and constraint approximation problems. Section~\ref{sec:grid} describes the grid generation methods, including the novel trajectory-based approach in non-uniform sampling of the input/state pairs. Section~\ref{sec:approx} defines the approximation problems. Section~\ref{sec:sim} presents the hybridization framework for model and constraint approximation using the generated grids and the validation results of the said approximation problems. Section~\ref{sec:conc} summarizes the hybridization framework, findings, and outlook for implementation and future work. This paper is Part I of a two-part publication entitled ``Efficient MPC for Emergency Evasive Maneuvers"; the application and analysis of the presented hybridization framework is discussed in detail in the second part: ``Efficient MPC for Emergency Evasive Maneuvers: Part II, Comparative Assessment for Hybrid Control".


\section{Background}\label{sec:back}

Consider a given nonlinear system, either in continuous-time $\dot{x} = F(x,u)$ or in discrete-time $x^+ = F(x,u)$ where $x \in \mathbb{R}^n$ and $u \in \mathbb{R}^m$ respectively represent the state and input vectors, and the domain of $F$ is denoted by $(x,u) \in \mathcal{D} \subseteq \mathbb{R}^{m+n}$. In many physics-based applications, the model $F$ is valid over a region $\mathcal{C} \subseteq \mathcal{D}$ defined by
\[\mathcal{C} \coloneqq \{ (x,u) \in \mathcal{D} \; \vert \; 0 \leqslant G(x,u) \leqslant 1\},\]
which collects a set of physics-based constraints\footnote{We use the normalized constraint formulation $0 \leqslant G \leqslant 1$ instead of the generic form $G \leqslant 0$ to avoid numerical issues in solving the approximation/control optimization problems.}. For instance, most typical vehicle models in the literature are no longer valid if e.g., the vehicle is rolling over. Here we aim at approximating both the nonlinear model $F$ and the nonlinear, non-convex set $\mathcal{C}$. Therefore, we need to hybridize both $F$ and $\mathcal{C}$. Both approximation problems can essentially be expressed as the minimization of the approximation error over their respective domains. The approximation error, as well as the domain, are different for each problem, as discussed hereafter.

\subsection{Model Approximation}

The system $F$ is approximated by a hybrid formulation $f$ via solving the nonlinear optimization problem 
\begin{align}
	\min_{\mathcal{A}} \int\limits_{\mathcal{C}} \dfrac{\left\Vert F(x,u) - f(x,u) \right\Vert_2}{\left\Vert F(x,u) \right\Vert_2 + \epsilon_0} \; d (x,u),
	\label{eq:fapprox}
\end{align}
where $\mathcal{A}$ represents the decision variables used to define $f$. The positive value $\epsilon_0>0$ added to the denominator is to avoid division by very small values for $\Vert F(x,u) \Vert_2 \approx 0$. Note that the domain in the model approximation problem is $\mathcal{C}$. 

\subsection{Constraint Approximation}

With the nonlinear, non-convex constraints given as $0 \leqslant G(x,u)\leqslant 1$, we approximate the feasible region $\mathcal{C}$ by a union of convex subregions $\mathcal{R}$. 

This approximation problem can be formulated in two ways: region-based and boundary-based. In the region-based approach, we minimize the misclassification error via solving the following optimization problem
\begin{align}
	\min_{\nu} \; \; \gamma_{\rm{c}} \dfrac{\mathcal{V} \{ \mathcal{C}\setminus \mathcal{R} \}}{\mathcal{V} \{ \mathcal{C}\}} +  (1-\gamma_{\rm{c}})  \dfrac{\mathcal{V} \{ \mathcal{R} \setminus \mathcal{C} \}}{\mathcal{V} \{ \mathcal{D} \setminus  \mathcal{C}\}},
	\label{eq:capprox}
\end{align}
where $\nu$ represents the decision variables used to define $\mathcal{R}$, the operator $\mathcal{V}$ gives the size or ``volume'' of the region, and $\gamma_{\rm{c}} \in [0,1]$ is a tuning parameter to adjust the relative penalization weight for the misclassification errors regarding inclusion error $\mathcal{C}\setminus \mathcal{R}$, i.e., failing to cover the feasible region, and the violation error $\mathcal{R} \setminus \mathcal{C}$ which corresponds to violating the constraints.

In the boundary-based approach, we approximate the boundary $G$ by a hybrid function $g$ and minimize the boundary-approximation error similar to (\ref{eq:fapprox}) via solving the optimization problem
\begin{align}
	\min_{\nu} \int\limits_{\mathcal{D}} \dfrac{\left\vert G(x,u) - g(x,u) \right\vert}{\left\vert G(x,u) \right\vert + \epsilon_0} \; d (x,u).
	\label{eq:gapprox}
\end{align}
with $\epsilon_0 > 0$. Note that as $G$ is a scalar function, the 2-norm is replaced by the absolute value here.

\begin{rem}
	The proposed ideas also apply in case of more inequalities e.g., 
	\[0 \leqslant G_i (x,u) \leqslant 1, \qquad \text{for } i \in \{1,2,\dots,N\},\]
	by simply formulating $G(x,u)$ as 
	\[G(x,u) = \max_{i \in \{1,2,\dots,N\}}\{G_i(x,u)\}.\]
	Another possibility is to approximate each $G_i$ independently; however, this may lead to redundant approximations of boundaries or parts of $G_i$ that do not belong to the overall boundary feasible region.
\end{rem}

\subsection{Relation to the State-of-the-Art}

The nonlinear non-convex constraints arise from the physics-based limitations of the system. Therefore,
\begin{itemize}
	\item the physics-based nature of the constraints results in a connected feasible region,
	\item the highly-nonlinear (boundary of the) constraints limits the analytical investigation of ``attainability"\footnote{Attainability of a point means that there exists an input such that the point is obtained by the system dynamics.} or optimality,
	\item the approximation approach is intended to be used within a hybridization benchmark, which means the method should be applicable for systems of higher degree and/or with high-dimensional feasible regions,
	\item the constraint violation is evaluated by ensuring that the solution lies within any of the subregions, which means overlapping subregions are acceptable,
	\item in light of improving the computational efficiency, it is desired to have a minimal approximation of the constraints, i.e.,\ approximating the non-convex feasible region with a union of fewer number of subregions is desired as well as an accurate coverage of the whole region, which leads to the need for
	\item a systematic approach to cover the non-convex feasible region by a union of convex subregions that allows balancing the violation vs.\ coverage of the approximation close to the constraint boundaries.
\end{itemize}

Considering the aforementioned features, the applicability of state-of-the-art methods based on convex-hull generation~\cite{Zhang2016} is limited for the current case as input-state spaces for complex vehicle models exceed four dimensions and a systematic division of the feasible region is not computationally efficient in terms of memory usage and speed for our desired accuracy. To compare our constraint approximation approach, we consider two state-of-the-art methods that share the most common elements with the aforementioned considerations in their respective problems.

The first method is from~\cite{Yao1996}, where a non-convex region is covered by a number of ellipsoids. There, an optimization problem is solved to minimize the misclassification error due to the region approximation where the center and radii of the ellipsoids are the decision variables. We refer to this approach as non-parametric elliptical learning, which is equivalent to region-based approximation of the constraints by a union of ellipsoids. Our constraint approximation framework can be seen an extension and generalization of this approach by investigating boundary-based vs.\ region-based approximations and polytopic vs.\ ellipsoidal definition of the subregions.

The second method is from~\cite{Duhr2022}, where the gripping limits of the vehicle are approximated by a convex intersection of second-order cone constraints. There, the constraints are formulated using the system dynamics and the parameters of the combined formulation are fitted using experimental data. We refer to this approach as the convex envelope method, which is equivalent to a boundary-based approximation of the constraints by the intersection of multiple convex subregion. Since this method approximates the non-convex feasible region by a convex one, in Section~\ref{sec:sim} we will show its limitation in converging to an accurate approximation of the constraints in comparison with our proposed framework.

Since analytical closed-form solutions for (\ref{eq:fapprox})--(\ref{eq:gapprox}) do not exist, we propose solving them numerically\footnote{For instance, another approach to solving the aforementioned approximation problem is the Monte Carlo integration method.} by generating a grid of samples from their regarding domains $\mathcal{C}$ and $\mathcal{D}$, respectively denoted by $\mathcal{C}^\ast$ and $\mathcal{D}^\ast$. As the grid generation method influences the quality of the final fit, we provide various grid-generation methods for both approximation problems in the next section and examine the resulting fits in our results in Section~\ref{sec:sim}.

\section{Grid Generation}\label{sec:grid}

We use two main approaches to generate~$\mathcal{D}^\ast$: domain-based and trajectory-based. In the domain-based approach, both the input and state elements of the grid points are selected from the input/state domain~$\mathcal{D}$, regardless of the system's behavior. While a domain-based grid can have a good coverage of $\mathcal{D}$, it does not take into account the ``likelihood" of the points being visited in a simulation with respect to the system dynamics. The trajectory-based way of generating $\mathcal{D}^\ast$ tackles this issue by selecting the input elements of the grid points $u^\ast$ from $\mathcal{D}$, while assigning the state elements to the points from an $n_\text{step}$-step-ahead simulation of $F$ given $u^\ast$ as the input. As a result, the obtained $\mathcal{D}^\ast$ will have a higher density in regions of $\mathcal{D}$ where the input/state pairs have a higher likelihood of being attainable. 

Each of these two approaches can be implemented in two ways, giving rise to a total of four methods to generate~$\mathcal{D}^\ast$:
\begin{itemize}
	\item \textbf{Domain-based:} [\emph{points are directly sampled from $\mathcal{D}$}]
	\begin{itemize}
		\item \textbf{Uniform ($\mathcal{D}^\ast_U$, also referred to as U grid type):} the points are generated by picking $n_\text{samp}$ uniformly-spaced points along each axis in~$\mathcal{D}$.
		\item \textbf{Random ($\mathcal{D}^\ast_R$, also referred to as R grid type):} a total of $n_\text{rand}$ points are randomly selected from $\mathcal{D}$.
	\end{itemize}
	\item \textbf{Trajectory-based:} [\emph{$n_\text{sim}$ open-loop simulations with $n_\text{step}$ steps of $F$ are run using random inputs from $\mathcal{D}$}]
	\begin{itemize}
		\item \textbf{Steady-state ($\mathcal{D}^\ast_S$, also referred to as S grid type):} the initial state of each simulation is selected as the steady-state solution w.r.t. the initial input, i.e., it is assumed that each simulation starts from a steady state.
		\item \textbf{Randomly-initiated ($\mathcal{D}^\ast_T$, also referred to as T grid type):} the initial state of each simulation is randomly selected from $\mathcal{D}$.
	\end{itemize}
\end{itemize}

Algorithms~\ref{alg:dbgrid} and \ref{alg:tbgrid} respectively explain the domain-based and trajectory-based grid generation methods. The total number of grid points for each type denoted by $\mathcal{N}$ is 
\begin{align*}
	\mathcal{N} (\mathcal{D}^\ast_U) &= (n_\text{samp})^{m+n},\\
	\mathcal{N} (\mathcal{D}^\ast_R) &= n_\text{rand}, \\
	\mathcal{N} (\mathcal{D}^\ast_S) = \mathcal{N} (\mathcal{D}^\ast_T) &= n_\text{sim} \cdot n_\text{step}.
\end{align*}
\vspace{-0.5cm}
\begin{algorithm}
	\caption{Domain-based grid generation}\label{alg:dbgrid}
	\begin{algorithmic}	
		\Require $F, \; \mathcal{D}, \; n_\text{samp},\; n_\text{rand}, \; \text{type} \in \{\text{`U', 'R'}\}$
		\State $\mathcal{D}^\ast_\text{type} \gets \{\}$
		\If{type = `U'}
		\For{$k \in \{1, 2, \dots , m+n\}$}	
		\State $\mathcal{I}_k \gets \{\}$
		\Comment{\emph{$\mathcal{I}_k \coloneqq$ sample set}}
		\For{$i \in \{0,  \dfrac{1}{n_\text{samp} - 1} , \dots, 1\}$} 
		\State $\mathcal{I}_k \gets \mathcal{I}_k \cup \{\mathcal{D}_{(k)_\text{min}} + \left(\mathcal{D}_{(k)_\text{max}} - \mathcal{D}_{(k)_\text{min}}\right) \; \cdot i \}$
		\EndFor
		\EndFor
		\State $\mathcal{D}^\ast_\text{U} \gets \mathcal{I}_1 \times \mathcal{I}_2 \times \dots \times \mathcal{I}_{m+n}$ 
		\Comment{\emph{Cartesian product}}
		\ElsIf{type = `R'}
		\For{$k \in \{1, 2, \dots , n_\text{rand}\}$}
		\State $(x_k,u_k) \xleftarrow{\text{ random }} \mathcal{D}$
		\State $\mathcal{D}^\ast_\text{R} \gets \mathcal{D}^\ast_\text{R} \cup \{(x_k,u_k)\}$ 
		\EndFor
		\EndIf		
		\State \Return $\mathcal{D}^\ast_\text{type}$
	\end{algorithmic}
\end{algorithm}
\vspace{-0.5cm}
\begin{algorithm}
	\caption{Trajectory-based grid generation}\label{alg:tbgrid}
	\begin{algorithmic}	
		\Require $F, \; \mathcal{D}_x, \; \mathcal{D}_u, \; n_\text{sim},\; n_\text{step}, \; \text{type} \in \{\text{`S', 'T'}\}$	
		\State $\mathcal{D}^\ast_\text{type} \gets \{\}$
		\For{$s \in \{1, 2, \dots, n_\text{sim}\}$}	
		\State $u \xleftarrow{\text{ random }} \mathcal{D}_u$ \Comment{$\mathcal{D}_u  \coloneqq $ \emph{input domain}}
		\State $x \xleftarrow{\text{ random }} \mathcal{D}_x$ \Comment{$\mathcal{D}_x  \coloneqq $ \emph{state domain}}
		\If{type = `S'}
		\State $x_1 \xleftarrow{\text{ solve for $x$ }}  F(x,u_1) = 0$ \Comment{\emph{steady-state solution}}
		\EndIf
		\For{$k \in \{2, 3, \dots, n_\text{step}\}$}
		\State $x_k \gets x_{k-1} + F\left(x_{k-1}, u_{k-1}\right)$									
		\If{$(x_k,u_k) \notin \mathcal{D}_x \times \mathcal{D}_u$} 
		\State \textbf{break}	 \Comment{\emph{stop current simulation}}
		\EndIf				
		\State $\mathcal{D}^\ast_\text{type} \gets \mathcal{D}^\ast_\text{type} \cup \{(x_k,u_k)\}$
		\EndFor		
		\EndFor
		\State \Return $\mathcal{D}^\ast_\text{type}$
	\end{algorithmic}
\end{algorithm}

\begin{figure}[htbp]
	\begin{center}
		\begin{subfigure}{0.38\textwidth}
			\includegraphics[width=\textwidth]{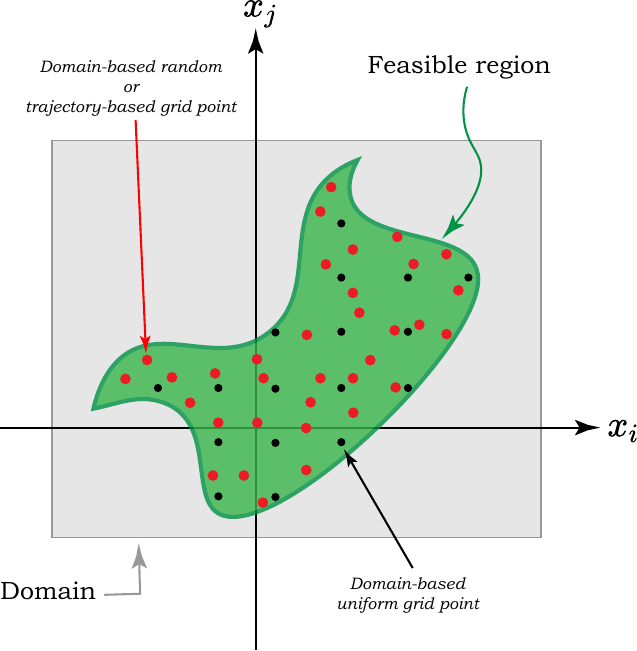}
			\subcaption{Grid generation for model approximation}
		\end{subfigure}
		\par\bigskip
		\begin{subfigure}{0.38\textwidth}
			\includegraphics[width=\textwidth]{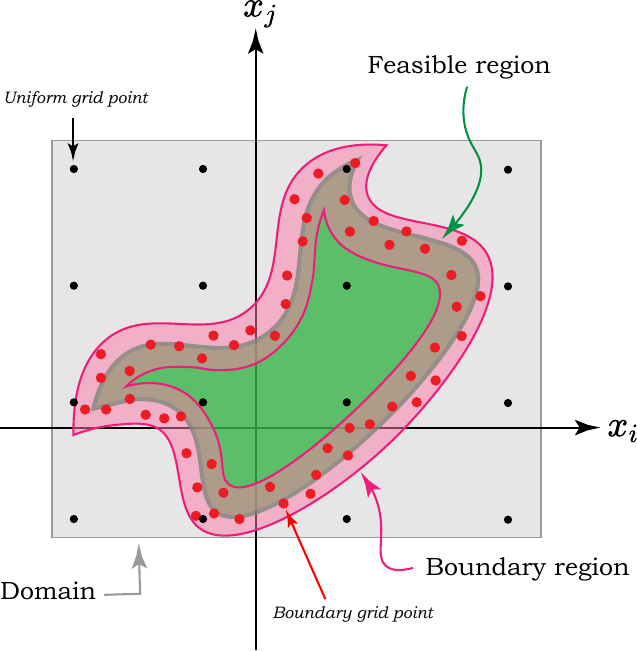}
			\subcaption{Grid generation for constraint approximation}
		\end{subfigure}
	\end{center}
	\caption{A schematic view of different implementations of the proposed grid-generation approaches for model and constraint approximation.}
	\label{fig:grids}
\end{figure}

The grid $\mathcal{D}^\ast$ plays the role of domain in the approximation problem. Therefore, it should be tailored to the objective of the problem itself. In this sense, Figure~\ref{fig:grids} shows a schematic view of the implementation of the proposed grid-generation approaches for both model and constraint approximation problems. 

For model approximation, the grid should be generated only from $\mathcal{C}$, as the points outside $\mathcal{C}$ are infeasible, which translates to zero likelihood of attainability. Therefore, while Algorithms~\ref{alg:dbgrid} and \ref{alg:tbgrid} are implemented on $\mathcal{D}$, only the samples from the feasible region should be kept. Then, the four resulting grids, $\mathcal{C}^\ast_U$, $\mathcal{C}^\ast_R$, $\mathcal{C}^\ast_S$, and $\mathcal{C}^\ast_T$ can be used to examine their efficacy.

Contrary to the model approximation problem, the points for constraint approximation should be distributed in the whole domain $\mathcal{D}$ to allow examining the approximation error. In addition, for constraint approximation, the areas close to the boundary of $\mathcal{C}$ are of more interest than the areas with higher likelihood of attainability. Therefore, while trajectory-based methods are useful for model approximation, to find the constraints, we are interested in using a domain-based grid with a higher density in the neighborhood of $G(x,u) = 0$. This grid can be obtained by combining a uniform grid $\mathcal{D}^\ast_U$ with a random grid $\mathcal{B}^\ast_R$ on the boundary region $\mathcal{B}$ where
\[\mathcal{B} \coloneqq \{ (x,u) \in \mathcal{D} \; \vert \; \vert G(x,u) \vert \leqslant \epsilon_b\}.\]
The resulting generated grid is $\mathcal{D}^\ast_U \cup \mathcal{B}^\ast_R$.

\begin{rem}
	To ensure that trajectory-based grids are generated by ``realistic" inputs, we impose a bound constraint on the random inputs as 
	\[\vert u^\ast (k+1) - u^\ast(k)\vert < \Delta_u^\ast.\]
	This can also account for the physical limitations of the actuators and be considered to be part of the physics-based constraints $\mathcal{C}$ and it is best selected based on data from real operation of the system.
\end{rem}

\begin{rem}
	Depending on the problem characteristics such as the system dynamics, domain, and the nature of the input/state signals, some points in the generated grids (except for the U grid type) can be very close to each other. To avoid these points from having larger importance than other points during approximation, Algorithms~\ref{alg:dbgrid} and \ref{alg:tbgrid} can further be refined by keeping only one point from each set of points that are closer to each other than a user-defined distance threshold.
\end{rem}

\section{Approximation Problem Formulation} \label{sec:approx}

\subsection{Model Approximation}

We approximate the nonlinear system $F$ by the MMPS function $f$ with the Kripfganz form~\cite{Kripfganz1987} as
\begin{equation}
	f (x,u) = \max_{p \in \{1, 2, \dots, P^+\} } \left\{\phi^+_p (x,u)\right\} - \max_{q \in \{1, 2, \dots, P^-\}} \left\{\phi^-_q (x,u)\right\},
	\label{eq:mmpsdef}
\end{equation}
where $P^+$ and $P^-$ are user-selected integers, and $\phi^+_p$, and $\phi^-_q$ are affine functions of $x$ and $u$, sometimes referred to as dynamic modes, and expressed as
\[\phi^+_p (x,u) =  A^+_p x + B^+_p u + H^+_p,\]
\[\phi^-_q (x,u) =  A^-_q x + B^-_q u + H^-_q.\]
We implement the MMPS approximation in the following fashion: each dimension of the  nonlinear function, i.e., each component of $F$, is approximated independently. Thus, $P^+$ and $P^-$, as well as the affine functions $\phi^+$ and $\phi^-$ are separately found for each component of $F$. Therefore, for brevity and without loss of generality, one can assume $F$ to be scalar in the remaining of this section.

For a fixed pair $(P^+,P^-)$ that corresponds to the number of affine terms in the first and second $\max$ operators in~(\ref{eq:mmpsdef}), we solve the nonlinear optimization problem (\ref{eq:fapprox}) subject to (\ref{eq:mmpsdef}) to find the optimal $\phi^+$ and $\phi^-$ functions where 
\begin{equation}
	\mathcal{A} = \left\{A^+_p, A^-_q, B^+_p, B^-_q , H^+_p, H^-_q\right\}_{p \in \{1, 2, \dots, P^+\}, q \in \{1, 2, \dots, P^-\}}.
	\label{eq:fdecvar}
\end{equation}

\begin{rem}
	To solve the nonlinear optimization problem in (\ref{eq:fapprox}), we generate a grid $\mathcal{C}^\ast$ of feasible samples from $\mathcal{D}$ as expressed in Section~\ref{sec:grid}, and minimize the objective function across $\mathcal{C}^\ast$.
	\label{rem:dtodast}
\end{rem}

\begin{rem}
	The Kripfganz form essentially expresses the function using $P^+ \cdot P^-$ hyperplanes as there are $P^+$ and $P^-$ affine functions in each max operator. Therefore, the hinging hyperplanes representing the local dynamics are obtained by subtraction of the affine functions $\phi^-$ from $\phi^+$ which means that the optimal $\mathcal{A}$ in (\ref{eq:fapprox}) would not be unique.
	\label{rem:addnorm}
\end{rem}

Considering Remarks~\ref{rem:dtodast} and \ref{rem:addnorm} and to avoid numerical problems, it is convenient to add a regularization term to (\ref{eq:fapprox}) by penalizing the 1-norm of the decision vector as 
\begin{align}
	\min_{\mathcal{A}} \int\limits_{\mathcal{C}^\ast} \dfrac{\left\vert F(x,u) - f(x,u) \right\vert}{\left\vert F(x,u) \right\vert + \epsilon_0} \; d (x,u) + \gamma_{\rm{m}} \Vert \mathcal{A} \Vert_1, && \text{s.t. } (\ref{eq:mmpsdef}),
	\label{eq:mmpsfindmodified}
\end{align}
where $\gamma_{\rm{m}} \in \mathbb{R}^+$ serves as a weighting coefficient to balance the penalization of the 1-norm of $\mathcal{A}$ with respect to the approximation error.

\subsection{Constraint Approximation}

\begin{figure}[htbp]
	\begin{center}
		\begin{subfigure}{0.35\textwidth}
			\includegraphics[width=\textwidth]{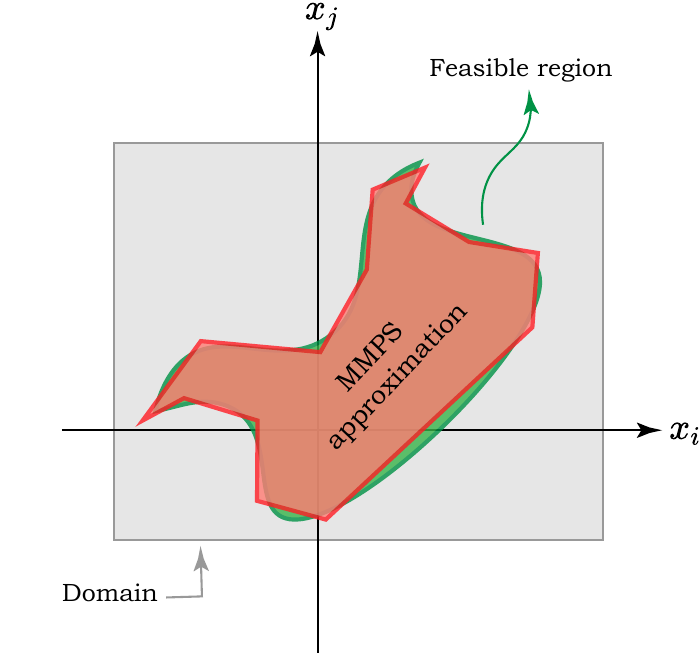}
			\subcaption{MMPS constraint approximation}
		\end{subfigure}
		\par\bigskip
		\begin{subfigure}{0.35\textwidth}
			\includegraphics[width=\textwidth]{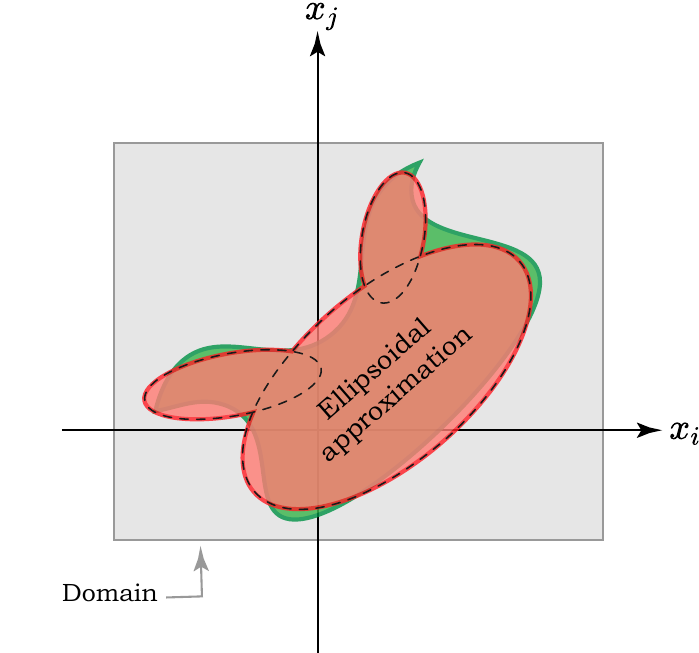}
			\subcaption{Ellipsoidal constraint approximation}
		\end{subfigure}
	\end{center}
	\caption{Illustration of MMPS and ellipsoidal approximation of the nonlinear constraints.}
	\label{fig:constapprox}
\end{figure}

We approximate the feasible region $\mathcal{C}$ by either a union of convex polytopes using the MMPS formalism, or by a union of ellipsoids. Figure~\ref{fig:constapprox} depicts both approaches to constraint approximation.

In the MMPS approach, a similar formulation to the MMPS model approximation problem is used: we approximate $G$ by an MMPS function $\gl$ of the Kripfganz form in (\ref{eq:mmpsdef}) with
\[\phi^+_p (x,u) =  C^+_p x + D^+_p u + I^+_p,\]
\[\phi^-_q (x,u) =  C^-_q x + D^-_q u + I^-_q.\]
The resulting feasible region $\Rl$ is then expressed as
\begin{equation}
	\Rl \coloneqq \{(x,u) \in \mathcal{D} \; | \; \gl(x,u) \leqslant 0\},
	\label{eq:mmpscdef}
\end{equation}
The MMPS approximation of the feasible region is then obtained  via solving either the region-based (\ref{eq:capprox}) or the boundary-based (\ref{eq:gapprox}) optimization problems subject to
\[\mathcal{R} = \Rl,\] 
and
\begin{equation}
	\nu = \left\{C^+_p, C^-_q, D^+_p, D^-_q , I^+_p, I^-_q\right\}_{p \in \{1, 2, \dots, P^+\}, q \in \{1, 2, \dots, P^-\}},
	\label{eq:gldecvar}
\end{equation}
where the matrices $C$, $D$, and $I$ represent the constraint-approximation counterparts of matrices $A$, $B$, and $H$ in (\ref{eq:fdecvar}) and $(P^+,P^-)$ stand for the respective number of affine terms.

The second way is to approximate the feasible region by a union of $n_{\rm{e}}$ ellipsoids 
\begin{align}
	\Rel \coloneqq  \left\{(x,u) \in \mathcal{D} \; \bigg| \;  \begin{pmatrix}
		x-x_{0_e} \\ u-u_{0_e}
	\end{pmatrix}^T Q_e \begin{pmatrix}
		x-x_{0_e} \\ u-u_{0_e}
	\end{pmatrix} \leqslant 1
	\right\},
	\label{eq:ellipdef}
\end{align}
with $Q_e$ being a positive definite matrix and $(x_0,u_0)$ representing the center coordinates of the ellipsoid. Note that this notation includes rotated ellipsoids as well. The approximated region $\Rq$ is
\begin{align}
	\Rq = \bigcup_{e=1}^{n_{\rm{e}}} \Rel \coloneqq \{(x,u) \in \mathcal{D} \; | \; \gq (x,u) \leqslant 0\},
	\label{eq:quniondef}
\end{align}
whose boundary can be expressed by
\begin{align}
	\gq (x,u) = \min_{e \in \{1, 2, \dots, n_{\rm{e}}\}} \left\{ 
	\begin{pmatrix}
		x-x_{0_e} \\ u-u_{0_e}
	\end{pmatrix}^T Q_e \begin{pmatrix}
		x-x_{0_e} \\ u-u_{0_e}
	\end{pmatrix} - 1 \right\}.
	\label{eq:ellipgdef}
\end{align}
The ellipsoidal approximation is found by solving either the region-based (\ref{eq:capprox}) or the boundary-based (\ref{eq:gapprox}) optimization problems subject to
\[\mathcal{R} = \Rq, \] 
and
\begin{equation}
	\nu = \left\{(x_{0_e},u_{0_e}),\; Q_e\right\}_{e \in \{1, 2, \dots, n_{\rm{e}}\}}.
	\label{eq:gqdecvar}
\end{equation}

\section{Model and Constraint Hybridization for Vehicle Control} \label{sec:sim}
In this section, the hybridization framework consisting of the model and constraint approximation approaches is implemented on a nonlinear single-track vehicle model with Dugoff tire forces and varying friction. First, the nonlinear system and physics-based constraints are described, then the training and validation grids are defined, which are next used for model and constraint approximation problems within the hybridization framework. The results are then discussed to evaluate the performance of the different approaches and analyzed for application in other nonlinear problems.

\subsection{Nonlinear System Descriptions}

\begin{figure}[htbp]
	\centering
	\includegraphics[width=0.45\textwidth]{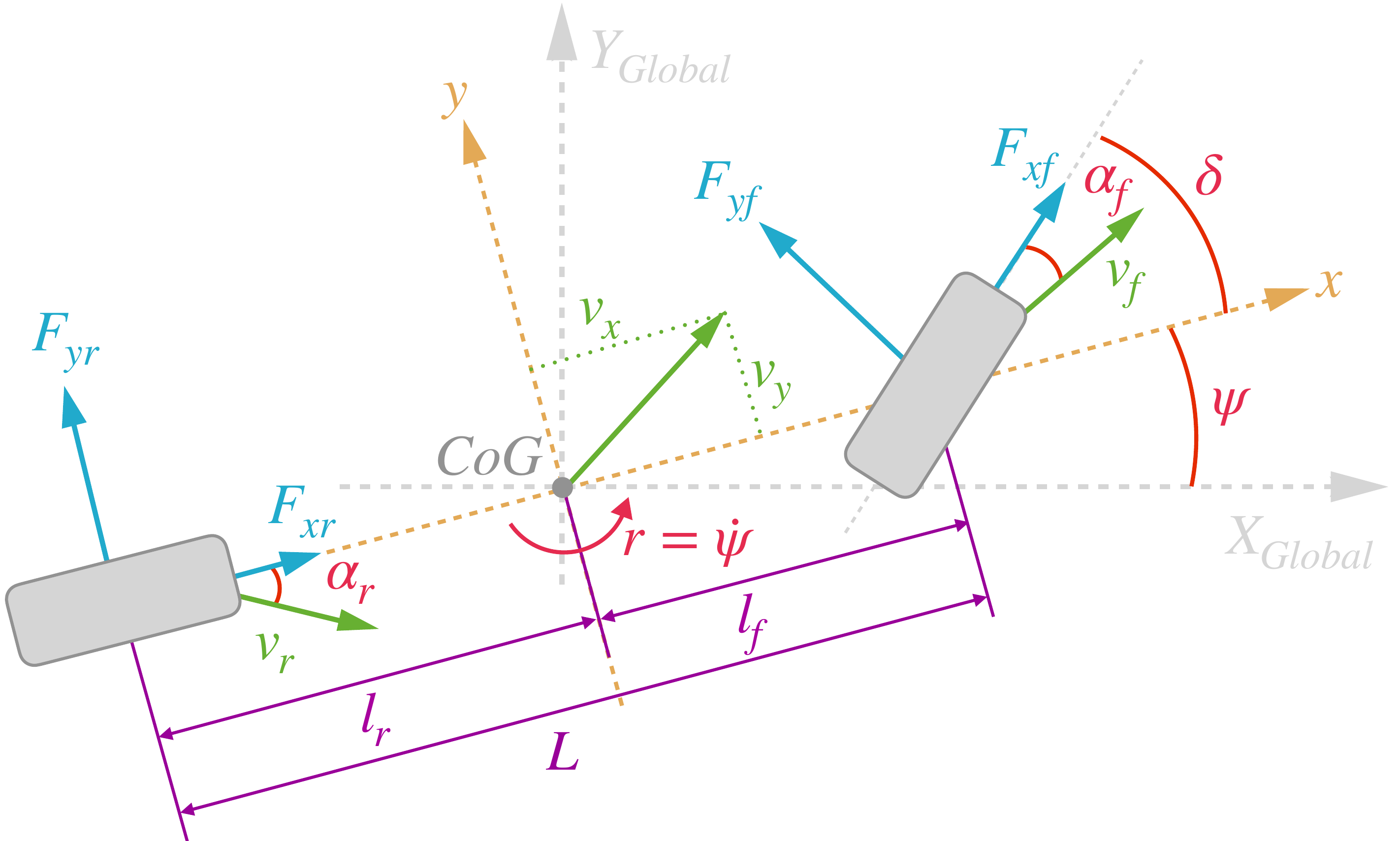}
	\caption{Configuration of the single-track vehicle model.}
	\label{fig:model}
\end{figure}

A single-track representation of the vehicle is shown in Fig.~\ref{fig:model}. With the system variables and parameters respectively defined in Tables~\ref{tab:vars} and \ref{tab:params}, the nonlinear vehicle model is described by the following equations~\cite{Chowdhri2021}:
\begin{equation}
	\dot{v}_x = \frac{1}{m} \left[F_{x\rm{f}} \cos{\delta} - F_{y\rm{f}} \sin{\delta} + F_{x\rm{r}}\right]+v_{y}r,
	\label{eq:vxnonlin}
\end{equation}
\begin{equation}
	\dot{v}_y = \frac{1}{m} \left[F_{x\rm{f}} \sin{\delta} + F_{y\rm{f}} \cos{\delta} + F_{y\rm{r}}\right]-v_{x}r,
	\label{eq:vynonlin}
\end{equation}
\begin{equation}
	\dot{r} = \frac{1}{I_{zz}} \left[F_{x\rm{f}} \sin{\delta} \; l_\text{f} + F_{y\rm{f}} \cos{\delta} \; l_\text{f} - F_{y\rm{r}} \;l_\text{r} \right],
	\label{eq:rnonlin}
\end{equation}
and the lateral forces are given by the Dugoff model
\[ F_{ya} = \dfrac{C_{\alpha_{a}}}{1-\kappa_{a}} f_\lambda(\lambda^w_{a}) \alpha_{a},\]
with $a \in \{\rm{f}, \rm{r}\}$ where $\mu_a$ is the varying friction coefficient, and $\lambda^w_{a}$ and $f_\lambda$ are the weighting coefficient and function, defined as
\[\mu_a = \mu_0 \left(1-e_r v_x \sqrt{\kappa_{a}^2 +\tan^2{{\alpha_{a}}}}\right),\]
\[\lambda^w_{a} = \frac{\mu_a F_{z{a}} (1-\kappa_{a})}{2 \sqrt{(C_{\kappa_{a}} \kappa_{a})^2+(C_{\alpha_{a}} \tan{\alpha_{a}})^2}},\]
\[f_\lambda(\lambda^w_a) =    
\begin{cases} 
	\lambda^w_{a}(2-\lambda^w_{a}) & \lambda^w_{a} < 1 \\
	1 & \lambda^w_{a} \geq 1
\end{cases}.\]

\begin{table}[htbp]
	\caption{System variables}
	\begin{center}
		\begin{tabular}{c|c|c|c}
			\hline
			\textbf{Var.} & \textbf{Definition} & \textbf{Unit} & \textbf{Bounds}\\
			\hline
			$v_x$ & Longitudinal velocity & m/s & [5, 50]\\
			$v_y$ & Lateral velocity & m/s & [-10, 10]\\
			$\psi$ & Yaw angle & rad & -- \\
			$r$ & Yaw rate & rad/s & [-0.6, 0.6]\\
			$\delta$ & Steering angle (road)& rad & [-0.5, 0.5]\\
			$F_{x\rm{f}}$ & Longitudinal force on the front axis & N & [-5000, 0]\\
			$F_{x\rm{r}}$ & Longitudinal force on the rear axis & N & [-5000, 5000]\\
			$F_{y\rm{f}}$ & Lateral force on the front axis & N & --\\
			$F_{y\rm{r}}$ & Lateral force on the rear axis & N & --\\
			$F_{z\rm{f}}$ & Normal load on the front axis & N & --\\
			$F_{z\rm{r}}$ & Normal load on the rear axis & N & --\\
			$\alpha_{\rm{f}}$ & Front slip angle & rad & --\\
			$\alpha_{\rm{r}}$ & Rear slip angle & rad & --\\
			$\kappa_{\rm{f}}$ & Front slip ratio & -- & --\\
			$\kappa_{\rm{r}}$ & Rear slip ratio & -- & --\\
			$\mu_{\rm{f}}$ & Friction coefficient on the front tire & -- & --\\
			$\mu_{\rm{r}}$ & Friction coefficient on the rear tire & -- & --\\
			\hline
			$x$ & State vector $\coloneqq \begin{bmatrix}
				v_x & v_y & r
			\end{bmatrix}^T$ & -- & -- \\
			$u$ & Input vector $\coloneqq \begin{bmatrix}
				F_{x\rm{f}} & F_{x\rm{r}} & \delta
			\end{bmatrix}^T$ & -- & -- \\
			\hline
		\end{tabular}
		\label{tab:vars}
	\end{center}
\end{table}

\begin{table}[htbp]
	\caption{System parameters$^\ast$}
	\begin{center}
		\begin{tabular}{c|c|c|c}
			\hline
			\textbf{Par.} & \textbf{Definition} & \textbf{Value} & \textbf{Unit}\\
			\hline
			$m$ & Vehicle mass & 1970 & kg\\
			$I_{zz}$ & Inertia moment about z-axis & 3498& kg/m$^2$\\
			$l_\text{f}$ & CoG$^{\ast\ast}$ to front axis distance& 1.4778 & m\\
			$l_\text{r}$ & CoG to rear axis distance & 1.4102 & m\\
			$C_{\alpha_\text{f}}$ & Front cornering stiffness &126784 & N \\
			$C_{\alpha_\text{r}}$ & Rear cornering stiffness & 213983 & N\\
			$C_{\kappa_\text{f}}$ & Front longitudinal stiffness & 315000 & N \\
			$C_{\kappa_\text{r}}$ & Rear longitudinal stiffness & 286700 & N\\
			$\mu_0$ & Zero-velocity friction & 1.076 & --\\
			$e_r$ & Friction slope & 0.01 & --\\
			\hline
			\multicolumn{4}{l}{$^{\ast}$These values correspond to the IPG CarMaker BMW vehicle model}\\
			\multicolumn{4}{l}{$^{\ast\ast}$Center of Gravity}
		\end{tabular}
		\label{tab:params}
	\end{center}
\end{table}

Table~\ref{tab:vars} also shows the bounds we impose on state and input vectors for grid generation. The feasible region is defined by two other physics-based constraints: 
\begin{enumerate}
	\item the working limits of the vehicle (known as the g-g diagram constraint~\cite{Chowdhri2021}) should be satisfied to allow derivation of the dynamics equation in (\ref{eq:vxnonlin}) to (\ref{eq:rnonlin}); this entails
	\begin{equation}
		\left(\dot{v}_x - v_y r\right)^2 + \left(\dot{v}_y + v_x r\right)^2 \leqslant (\min_{a \in \{\rm{f},\rm{r}\}} \{\mu_a g\})^2,
		\label{eq:gg}
	\end{equation}
	\item the tires can provide forces up to their saturation limit, known as the Kamm circle constraint~\cite{Chowdhri2021}, which means
	\begin{equation}
		F_{xa}^2 + F_{ya}^2 \leqslant (\mu_a F_{za})^2, \quad a \in \{\rm{f}, \rm{r}\}.
		\label{eq:kamm}
	\end{equation}		
\end{enumerate}
Therefore, the feasible region $\mathcal{C}$ can be expressed as
\[\mathcal{C} \coloneqq \left\{(x, u) \in \mathcal{D} \; \vert \; (\ref{eq:gg}) \; , \; (\ref{eq:kamm}) \right\}.\]

\subsection{Grid Definition and Coverage}

Table~\ref{tab:tgrids} shows the grid properties for the model and constraint approximation problems. For the model, all four U, R, S, and T grid types are used for training and later validated on a finer U, R, S, T grid type, respectively, plus C grid type, that is a grid that combines all of them. For the constraint approximation, only one combined grid consisting of the union U and R grids is used for training and the approximations are validated on a finer and more extended combined grid. 

\begin{figure}[htb]
	\centering
	\includegraphics[width=0.4\textwidth]{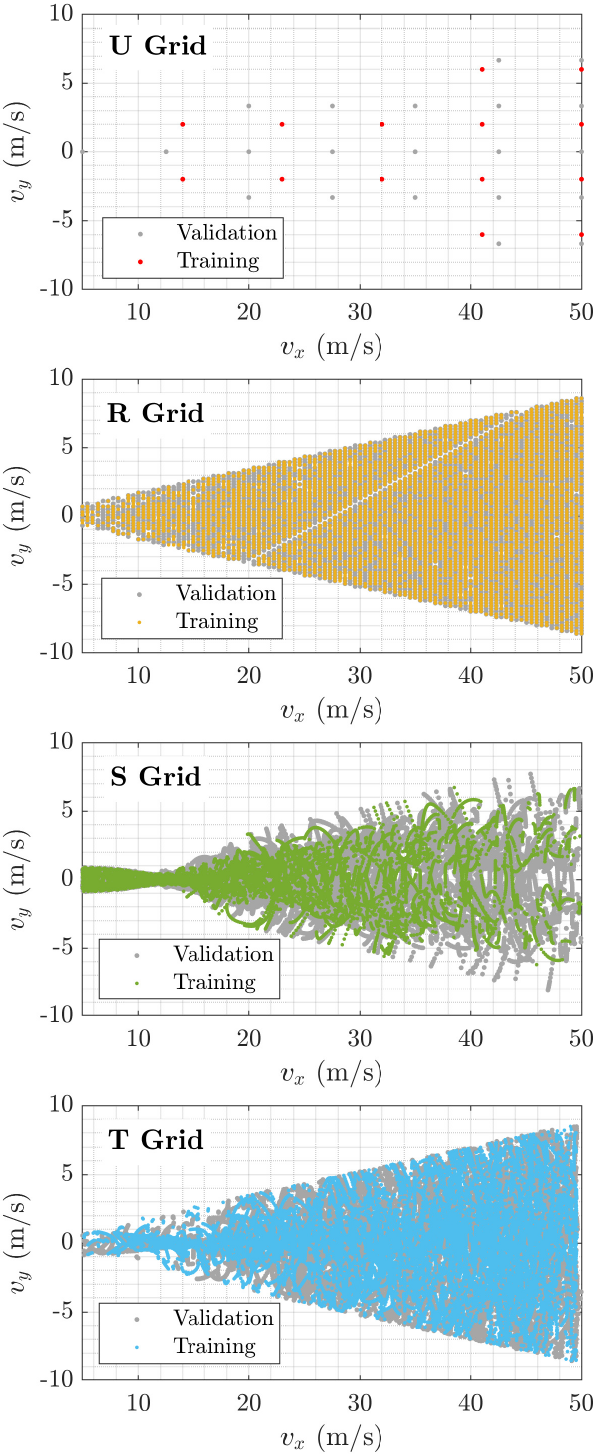}
	\caption{Location of training and validation grid points in the $v_y-v_x$ domain for different grid-generation approaches in model approximation}
	\label{fig:mgrids}
\end{figure}

For a visual comparison of the grid-point distribution for different types, we have plotted the coverage of the model approximation training and validation grids in the velocity domain ($v_x$-$v_y$) in Fig.~\ref{fig:mgrids}. While the grids have a similar total number of points, the density of the points among different grid types varies significantly as follows:
\begin{enumerate}
	\item The domain-based grids cover $\mathcal{C}$ with a uniform density compared to the trajectory-based grids.
	\item Compared to its random counterpart, the U grid represents a sparser distribution in the velocity domain, which stems from the fact that representation of all the possible combinations of input/state pairs on lower-dimensional sub-spaces of $\mathcal{C}$ projects many points on the exact same location in the viewed plane. 
	\item Between the trajectory-based grids, the randomly-initiated type (T) gives a better coverage of $\mathcal{C}$. Contrarily, the S grid favors the regions of $\mathcal{C}$ where the states are attainable from a steady-state solution within a bounded number of steps, which explains the high density of points in low-speed region and the loose coverage of high-speed regions with zero lateral velocity.
\end{enumerate}
\begin{table}[htb]		
	\caption{Properties of the grid used in the approximation problems (training and validation grids)}
	\label{tab:tgrids}
	\begin{center}
		\begin{tabular}{c c c c c}
			\toprule
			\multicolumn{5}{c}{\emph{Training Grids for Model Approximation}} \\
			\midrule
			\textbf{Type} & \textbf{Domain} & \textbf{Properties} & \textbf{No. Points} & \textbf{Feasible}\\
			\midrule
			\textbf{U} & $\mathcal{C}$ & $n_\text{samp}$ = 6 & $\approx 7,000$ & 100\%\\
			\textbf{R}  & $\mathcal{C}$ & $n_\text{rand}$ = 7000 & $\approx 7,000$ & 100\%\\
			\textbf{S}  & $\mathcal{C}$ & $n_\text{sim}$ = 500, $n_\text{step}$ = 1000 & $\approx 7,000$ & 100\%\\
			\textbf{T}  & $\mathcal{C}$ & $n_\text{sim}$ = 300, $n_\text{step}$ = 1000 & $\approx 7,000$ & 100\%\\
			\midrule[0.75pt]
			\multicolumn{5}{c}{\emph{Validation Grids for Model Approximation}} \\
			\midrule
			\textbf{Type} & \textbf{Domain} & \textbf{Properties} & \textbf{No. Points} & \textbf{Feasible}\\
			\midrule			
			\textbf{U} & $\mathcal{C}$ & $n_\text{samp}$ = 7 & $\approx 21,000$ & 100\%\\
			\textbf{R} & $\mathcal{C}$ & $n_\text{rand}$ = 21,000 & $\approx 21,000$ & 100\%\\			
			\textbf{S} & $\mathcal{C}$ & $n_\text{sim}$ = 3000, $n_\text{step}$ = 1000 & $\approx 21,000$ & 100\%\\			
			\textbf{T} & $\mathcal{C}$ & $n_\text{sim}$ = 1200, $n_\text{step}$ = 1000 & $\approx 21,000$& 100\%\\
			\textbf{C}  & $\mathcal{C}$ & combining all the above & $\approx 84,000$ & 100\%\\
			\toprule
			\multicolumn{5}{c}{\emph{Training Grids for Constraint Approximation}} \\
			\midrule
			\textbf{Type} & \textbf{Domain} & \textbf{Properties} & \textbf{No. Points} & \textbf{Feasible}\\
			\midrule
			\textbf{U} & $\mathcal{D}$ & $n_\text{samp}$ = 5 & $\approx 15,000$ & 68\% \\
			\textbf{R} & $\mathcal{B}$ & $n_\text{rand}$ = 15,000, $\epsilon_\text{b}$ = 0.1 & $\approx 15,000$ & 41\%\\
			\textbf{C} & $\mathcal{D}$ & combining all the above & $\approx 30,000$ & 55\% \\
			\midrule[0.75pt]
			\multicolumn{5}{c}{\emph{Validation Grids for Constraint Approximation}}\\
			\midrule
			\textbf{Type} & \textbf{Domain} & \textbf{Properties} & \textbf{No. Points} & \textbf{Feasible}\\
			\midrule			
			\textbf{U} & $\mathcal{D}$ & $n_\text{samp}$ = 6 & $\approx 47,000$ & 68\% \\
			\textbf{R} & $\mathcal{B}$ & $n_\text{rand}$ = 45,000, $\epsilon_\text{b}$ = 0.2 & $\approx 45,000$ & 56\% \\
			\textbf{C} & $\mathcal{D}$ & combining all the above & $\approx 92,000$ & 62\% \\
			\bottomrule
		\end{tabular}
	\end{center}
\end{table}

\begin{figure}[htb]
	\centering
	\includegraphics[width=0.4\textwidth]{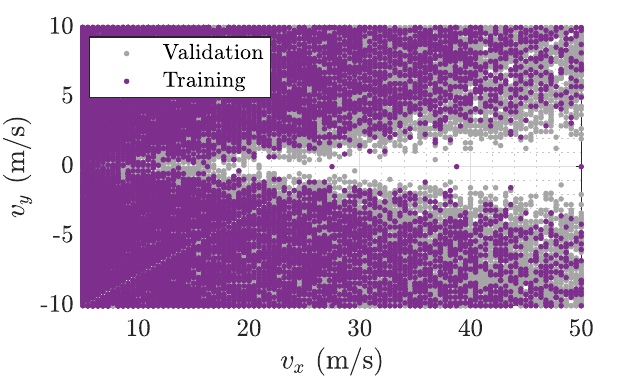}
	\caption{Location of the training and validation combined grid points in the $v_y-v_x$ domain for constraint approximation}
	\label{fig:cgrids}
\end{figure}

The constraint approximation grids in the velocity domain are shown in  Fig.~\ref{fig:cgrids}. Besides generating more grid points in the validation grids, the width $\epsilon_b$ of its boundary region is selected twice as large as for the training one, which increases the relative density of the grid points in the high-speed region as visible in Fig.~\ref{fig:cgrids}. Moreover, both grids have 50-60\% of their points in the feasible region, which is a reasonable ratio for a fair comparison.

\subsection{Model Approximation Results}

Using the four model training grids in Table~\ref{tab:tgrids}, we approximate the dynamics of the three states independently by Kripfganz MMPS functions with $(P^+,P^-)$ with $P^+ , P^- \in \{1, 2, \dots 8\}$. Since the approximated model will eventually be discretized before being incorporated in the MPC formulation, we already use a discretized form of the dynamics $\dot{x}$ in (\ref{eq:vxnonlin}) to (\ref{eq:rnonlin}) for approximation as
\[x(k+1) = \Delta x(k) + x(k).\] 
Here, $\Delta x(k)$ is approximated instead of $x(k+1)$ for two reasons: first, the assumptions and the approximation procedure remains valid by switching from $\dot{x}$ to $\Delta x$, and second, in cases such as $v_x$ where the state values are of a significantly larger order of magnitude compared to their rates of change, approximating $\Delta x$ leads to a more numerically-stable representation of the error. 

We solved the optimization problem (\ref{eq:mmpsfindmodified}) for every fixed pair of $(P^+,P^-)$ by \textsc{Matlab}'s nonlinear least squares optimizer, \texttt{lsqnonlin}, using the trust-region-reflective algorithm. This optimizer further exploits the structure of the nonlinear problem by approximating the Gauss-Newton direction through minimizing the 2-norm of the function deviation in the next step. The problem is then solved for 1000 initial random guesses to provide sufficient accuracy without excessive computational effort, among which we select the lowest objective value as the optimal solution. The codes for grid generation and hybrid approximations are available from our published hybridization toolbox~\cite{HybridCode}.

\begin{figure}[htbp]
	\begin{center}
		\begin{subfigure}{0.41\textwidth}
			\includegraphics[width=\textwidth]{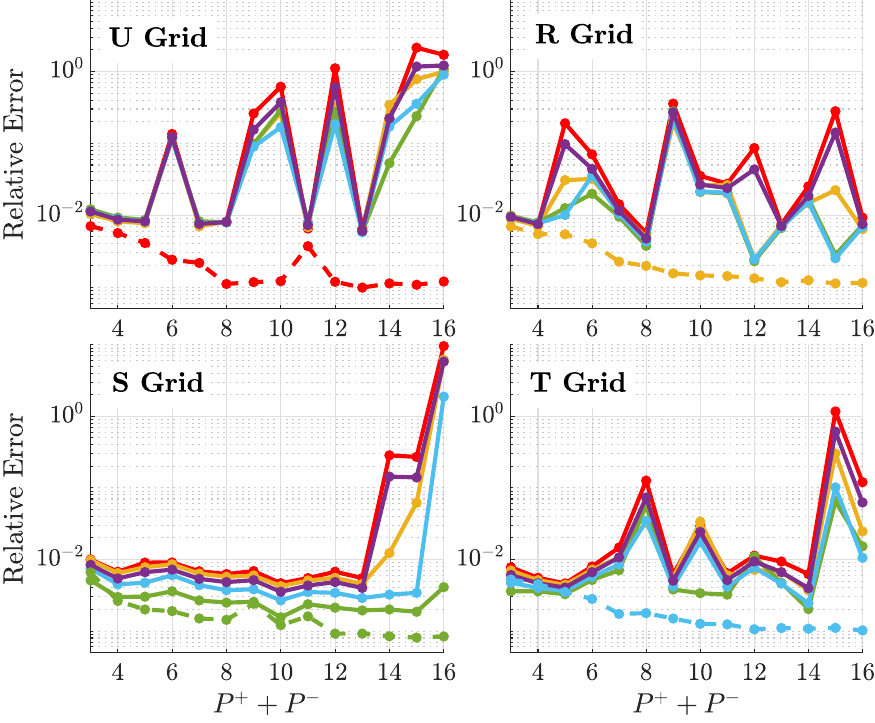}
			\subcaption{$\Delta v_x$}
		\end{subfigure}
		\par\bigskip
		\begin{subfigure}{0.41\textwidth}
			\includegraphics[width=\textwidth]{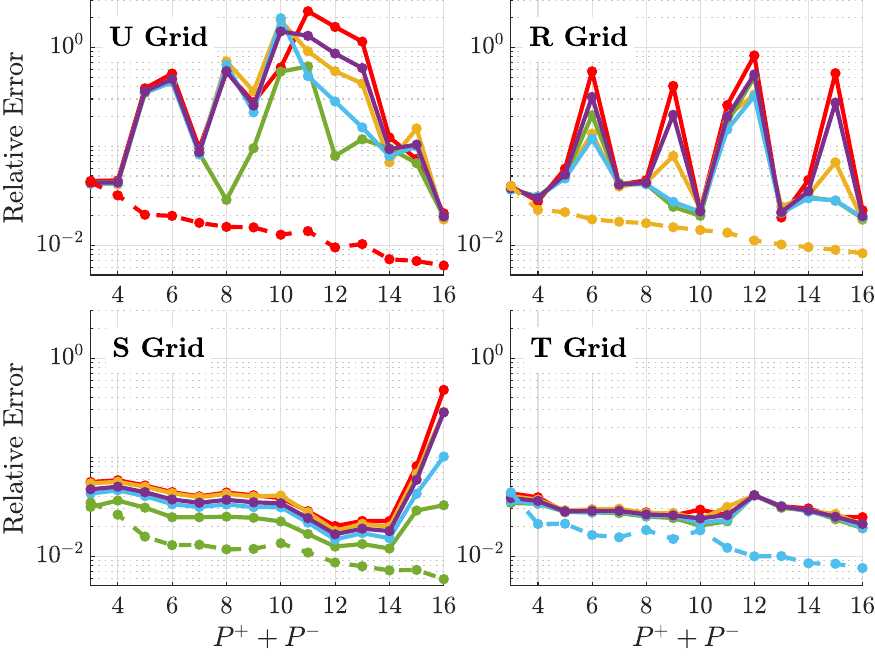}
			\subcaption{$\Delta v_y$}
		\end{subfigure}
		\par\bigskip
		\begin{subfigure}{0.41\textwidth}
			\includegraphics[width=\textwidth]{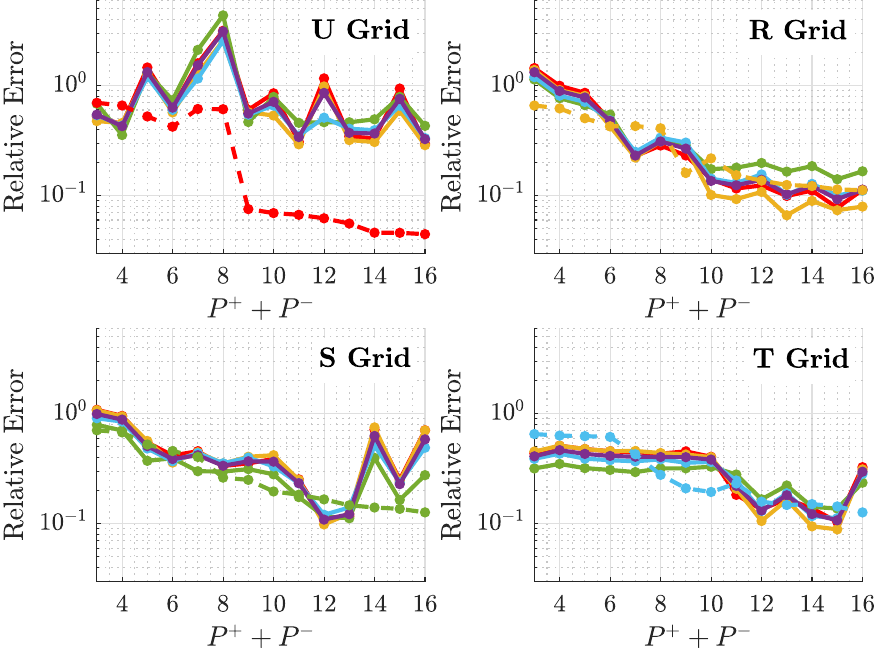}
			\subcaption{$\Delta r$}
		\end{subfigure}
		\par\bigskip
		\begin{subfigure}{0.13\textwidth}
			\includegraphics[width=\textwidth]{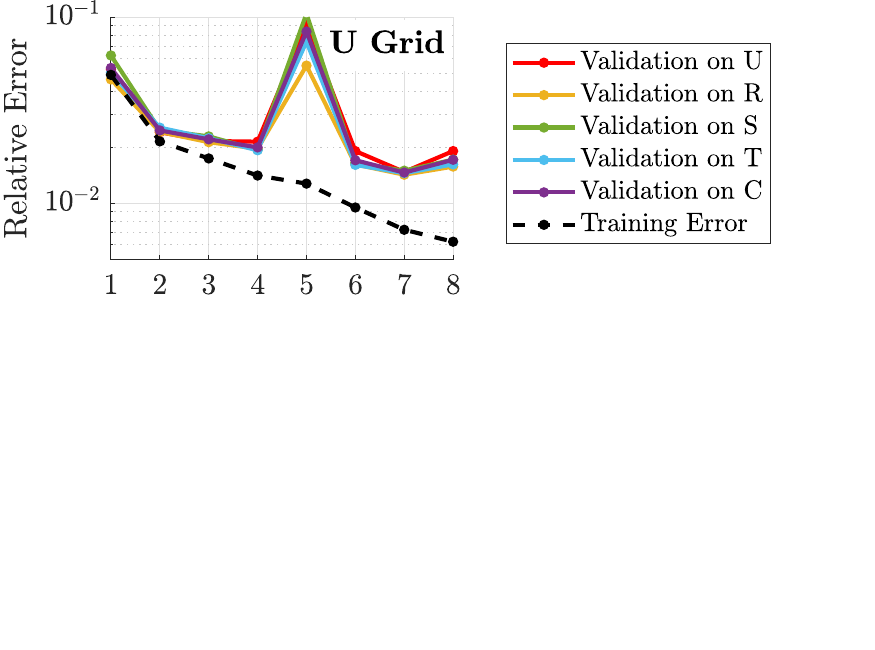}
			\subcaption{Plot legend}
		\end{subfigure}
	\end{center}
	\caption{Cross-validation of the MMPS approximations for different dynamics using four grid types. Since all the plots share the same legend, it is placed separately.}
	\label{fig:crossval}
\end{figure}

Fig.~\ref{fig:crossval} shows the training validation errors of the optimal solutions for $\Delta v_x$, $\Delta v_y$, and $\Delta r$ on model approximation validation grids in Table~\ref{tab:tgrids}. The lateral dynamics of the nonlinear model has a higher degree of nonlinearity, which explains the different error scales in the MMPS approximation. The plots are grouped based on the system and the type of the training grid to gain a better insight into the behavior of each grid and its effect on the accuracy of the approximation. 

Firstly, it is observed that U and R grids overfit for lower numbers of hyperplanes compared to their trajectory-based counterparts, which is represented by high oscillations after a certain degree of complexity in the approximation form. The S grid shows the lowest oscillatory behavior in validation results, which can indicate the inability of this grid in converging to an accurate fit due to its grid-point distribution with higher density in regions that are attainable from a steady-state solution of the system dynamics.

For $\Delta v_x$, U and R grids show overfitting behavior for \mbox{$P^++P^-\geqslant4$} modes and T grid overfits for \mbox{$P^++P^-\geqslant5$}. However, the S grid does not show overfitting until 13 modes with a lower validation error ($\approx 0.4\%$) compared to the other grid types ($\approx 0.8\%$). It is worth noting that the trajectory-based validation grids start overfitting for a much larger number of modes compared to the domain-based types.

For $\Delta v_y$, U and R grids again overfit at 4 modes, with 3\% and 2\% validation errors, respectively. The S grid overfits at 12 to 14 modes with reaching a validation error that is slightly above 1\%, and the T grid overfits at 11 modes with an error of 2\%. 

For $\Delta r$, U grid overfits at 4 modes and its validation error remains above 42\%. On the other hand, the R, S, and T grids reach their best fits at 12 to 15 modes, all with an error of about 9\%. 
The S grid, while having the lowest training error in most cases, has the highest offset between the validation and the training error. This could be due to the S grid needing more points to provide a more realistic training error. However, it should be noted that the steady-state-initiated method's ability to generate new ``distinct" points is limited; as Table~\ref{tab:tgrids} shows, to generate a validation grid three-times as large as the training one, the number of simulations needed to be multiplied by 6, which is not the case for its randomly-initiated counterpart, T. As the set of points attainable by a random input signal from a steady-state solution is limited, this difference is understandable. Nevertheless, this limitation is not restricting the S grid's ability to fit the model significantly (compared to e.g., the U grid). 

\begin{table}[htb]		
	\caption{Best validation fits for different grid types}
	\label{tab:mgoods}
	\begin{center}
		\begin{tabular}{c | c  c | c c | c c }
			\toprule
			Grid & \multicolumn{2}{c}{$\Delta v_x$} &
			\multicolumn{2}{c}{$\Delta v_y$} &
			\multicolumn{2}{c}{$\Delta r$}\\
			\cmidrule{2-7}
			type & $(P^+,P^-)$ & Error$^\ast$ & $(P^+,P^-)$ & Error$^\ast$ & $(P^+,P^-)$ & Error$^\ast$ \\
			\midrule
			\textbf{U} & (2,3) & 0.8\% & (2,2) & 4.3\% & (2,2) & 42.8\% \\
			\textbf{R} & (2,2) & 0.7\% & (2,2) & 3.0\% & (7,8) & 9.2\% \\
			\textbf{S} & (3,7) & 0.3\% & (6,8) & 1.8\% & (6,6) & 9.8\% \\
			\textbf{T} & (2,3) & 0.5\% & (6,3) & 2.6\% & (7,8) & 8.8\% \\
			\bottomrule
		\end{tabular}
	\end{center}
\footnotesize{$^\ast$ Relative validation error on the C grid}
\end{table}

\subsection{Constraint Approximation and Validation}

For constraint approximation, both training and validation steps  are done on the two constraint approximation C grids defined in Table~\ref{tab:tgrids}. The nonlinear constraints are approximated by either an intersection of second-order cones, which corresponds to the implementation of the convex envelope method from~\cite{Duhr2022}, or a union of convex subregions, which gives a non-convex approximation of the feasible region. Based on the formulation of the approximation problem, i.e., (\ref{eq:capprox}) or (\ref{eq:gapprox}), the approach is region- or boundary-based. The shape of the subregions is also either ellipsoidal or polytopic, where the latter is developed by an MMPS formulation of the nonlinear constraint. This leads to four methods of constraint approximation as shown in Table~\ref{tab:cgoods} where the best fits and their corresponding parameters as well as their approximation errors are presented. It should be noted that the region-based ellipsoidal approximation is a modified implementation of the non-parametric ellipsoidal learning method~\cite{Xu1996}.

\begin{figure}[htbp]
	\begin{center}
		\begin{subfigure}{0.49\textwidth}
			\includegraphics[width=\textwidth]{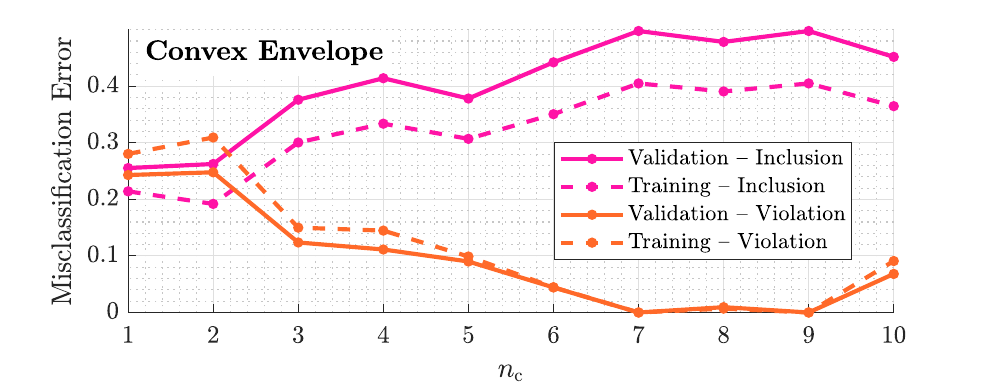}
			\subcaption{Intersection of convex subregions}
			\label{fig:constvala}
		\end{subfigure}
		\par\bigskip
		\begin{subfigure}{0.44\textwidth}
			\includegraphics[width=\textwidth]{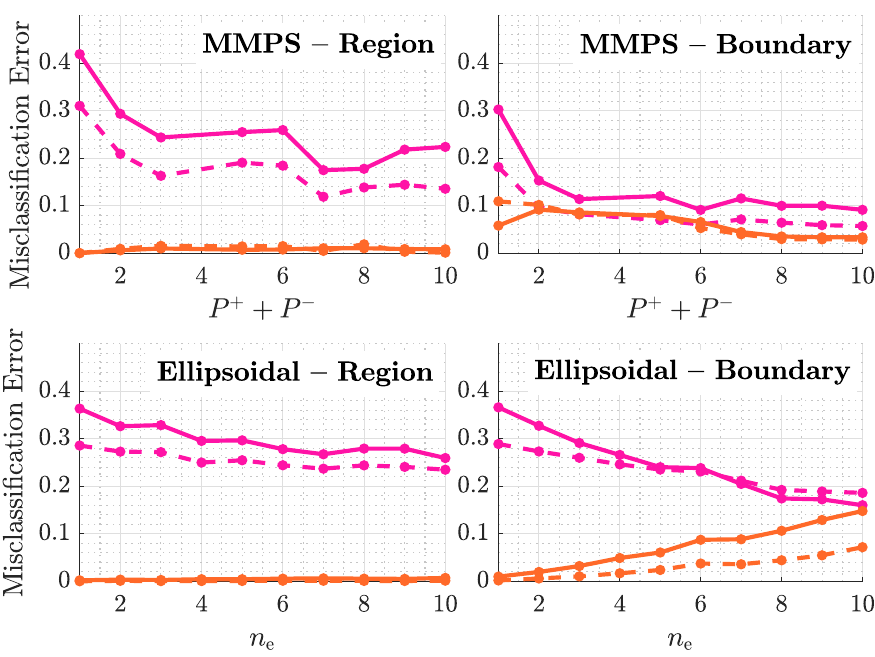}
			\subcaption{Union of convex subregions}
			\label{fig:constvalb}
		\end{subfigure}
	\end{center}
	\caption{Training and validation plots for different constraint approximation problems. As the axes share the same legend, it is only presented in the first one.}
	\label{fig:constval}
\end{figure}

Similar to model approximation, we solved the boundary-based optimization problems (\ref{eq:gapprox}) for every fixed pair of $(P^+,P^-)$ or $n_{\rm{e}}$ by \textsc{Matlab}'s nonlinear least squares optimizer, \texttt{lsqnonlin} for 1000 initial guesses (selected in a similar way as for the model approximation). However, the region-based approach results in a non-smooth optimization problem (\ref{eq:capprox}) which we solved using the particle swarm optimizer in \textsc{Matlab}, which does not require the problem to be differentiable. The swarm size was selected to be 10 times larger than the number of decision variables as a sufficiently large number for our experiments, and the problem was solved 1000 times for each case of $(P^+,P^-)$ or $n_{\rm{e}}$ and the best solution was kept as the optimal one. In addition, the convex envelope approach from~\cite{Duhr2022} where the boundary of the nonlinear constraints is approximated by an intersection of $n_{\rm{c}}$ second-order cone constraints is also implemented in the same fashion for different values of $n_{\rm{c}}$. Figure~\ref{fig:constval} shows the training and validation errors for different constraint approximation methods.

The convex envelope approach approximates the feasible region by a convex area that is the intersection of $n_{\rm{c}}$ second-order cone constraints. Therefore, for systems where the concavity measure, i.e., the difference between the feasible region and its convex hull, is significant compared to its size, this method converges to either high violation or inclusion misclassification errors, which is visible in the behavior of the training and validation plots in Fig.~\ref{fig:constvala}. Starting from one second-order cone constraint to approximate the feasible region with, this approach converges to an area covering about 25\% of the feasible and 25\% of the infeasible regions. Increasing the number of cone constraints to more than 3 leads to a significant improvement in the obtained fit. Nevertheless, the best convex envelope fit is obtained at $n_{\rm{c}} = 6$ with the inclusion and violation errors of 45\% and 5\% respectively, both of which are not acceptable as a proper fit. This shows that the method is converging to more accurate approximations of the largest convex subset of the feasible region, which is covering about 50\% of it.

The difference between the region- and boundary-based approaches is due the fact that in the region-based approximation (\ref{eq:capprox}), the inclusion and violation misclassification errors are penalized, while in the  boundary-based approximation (\ref{eq:gapprox}), the error in approximation of the distance to the boundary is minimized. This difference is more clear in the MMPS approximation plots where with one binary variable, the boundary is approximated by an affine function, i.e., a hyperplane. Problem (\ref{eq:gapprox}) then converges to a hyperplane with the lowest sum of distances from the nonlinear boundary. However, since the violation error is penalized more than the inclusion error with $\gamma_{\rm{c}} <0.5$, problem (\ref{eq:capprox}) converges to an empty set where the violation error is zero and the inclusion error is 1, giving the optimal misclassification error of $1-\gamma_{\rm{c}}$. In all the cases, it is observed that the region-based approximation converges to lower violation and higher inclusion errors due to the same reason.

MMPS approximation of the constraints via the region-based approach shows overfitting behavior after considering 6 binary variables. After 3 binary variables, the fits start oscillating between a more ``inclusive" approximation and a more ``violating" one. However, the best fit is obtained with 7 binary variables. Even by increasing this number, problem (\ref{eq:capprox}) keeps converging to the same misclassification error.

Boundary-based MMPS approximation reaches the best fit with 8 binary variables where again, adding more binary variables and increasing the complexity level of the fit does not change the inclusion and violation errors significantly and only minor oscillations between converging to a slightly more inclusive approximation or to a slightly more violating one are observed.

Ellipsoidal approximation of the feasible region generally converges to fits with lower accuracy compared to the MMPS approximation. In the region-based approximation, the training and validation errors stay at the same level with slight oscillations after $n_{\rm{e}} = 7$ with inclusion and violation misclassification errors of respectively 26.7\% and 0.6\%. In this sense, for the same number of integer variables, the ellipsoidal region-based approximation converges to a similar violation error but a 50\% higher inclusion error. The boundary-based ellipsoidal approximation on the other hand shows a different overfitting behavior where increasing the number of ellipsoidal subregions results in convergence to a better coverage at the expense of a significant increase in violation error. Therefore, the best fit should be selected before the point where the violation error exceeds a user-defined accepted threshold. Here we select $n_{\rm{e}} = 5$ since it is the last complexity before the violation error exceeds 6\%. Another observed pattern is the divergence of violation errors in training and validation, which mirrors the nature of the approximation approach: increasing the number of ellipsoids translates into generating more ellipsoidal subregions close to the boundary to minimize the distance-to-boundary sum. However, in the validation phase this leads to significantly higher violation errors as a result of the approximation overfitting to the training grid. 
 
\begin{table}[htb]		
	\caption{Best constraint approximation fits}
	\label{tab:cgoods}
	\begin{center}
		\begin{tabular}{c c | c | c | c}
			\toprule
			\multirow{2}{*}{\textbf{Subregions}} & \multirow{2}{*}{\textbf{Approach}} &
			\multirow{2}{*}{\textbf{Fit Parameters}} & \multicolumn{2}{c}{\textbf{Error}} \\
			& & & Inclusion & Violation \\
			\midrule
			\multicolumn{5}{c}{\emph{Intersection of convex subregions}~\cite{Duhr2022}}\\
			\midrule		
			Cone & Boundary & $n_c$ = 6 & 45.0\% & 5.0\% \\
			\midrule
			\multicolumn{5}{c}{\emph{Union of convex subregions}}\\
			\midrule
			MMPS & Region & $(P^+,P^-) = (5,2)$ & 17.5\% & 0.5\% \\
			MMPS & Boundary & $(P^+,P^-) = (4,4)$ & 9.9\% & 3.5\% \\
			\midrule
			Ellipsoidal & Region~\cite{Xu1996} & $n_{\rm{e}} = 7$ & 26.7\% & 0.6\% \\
			Ellipsoidal & Boundary & $n_{\rm{e}} = 5$ & 24.0\% & 6.0\% \\
			\bottomrule
		\end{tabular}
	\end{center}
\end{table}

\section{Conclusions and Outlook}\label{sec:conc} 

This paper has presented a hybridization framework for approximation nonlinear model and constraints. This framework serves as benchmark for formulating nonlinear MPC optimization problems using a hybrid systems formalism to improve computational efficiency and to ensure real-time implementation. The conclusions of the research in this paper with respect to its contributions, and the hybridization framework are summarized in the following subsections. The hybrid control comparison benchmark is discussed in detail in Part II of this publication. 

\subsection{Conclusions for Vehicle Control}

Introduction of the hybridization framework in this paper is a result of the following steps where the model and constraint approximation problems were defined by means of several novel descriptions of the approximation problem. First, for the model approximation, the Kripfganz MMPS form was used to approximate the nonlinear system to a user-defined error bound. Second, the nonlinear feasible region resulting from the physics-based constraints was approximated by a union of ellipsoids and polytopes via region- and boundary-based formulation of the approximation problem. Third, the model and constraint approximation problems were solved numerically across various grids types sampled from the input/state domain and their corresponding fit qualities in terms of accuracy and overfitting behavior were compared. Fourth, among the different grid types, two novel trajectory-based grid generation methods were introduced to structurally increase the density of the grid points in regions of the state domain with higher likelihood of the attainability by the system dynamics. This approach resulted in $15$-$60$\% reduction of the approximation error compared to its domain-based counterpart. Finally, the different grid generation and formulations of the approximation problems were analyzed to present a hybridization benchmark for improving the computational performance of the MPC problem for other applications of nonlinear MPC, as well as tracking control in emergency evasive maneuvers; this comparative assessment is explained in Part II.

\subsection{Generalized Hybridization Framework}

Our proposed hybridization framework can be implemented in other applications of nonlinear MPC to improve computational efficiency by considering the following guidelines:

\begin{enumerate}
	\item The model approximation problem should be solved by either an R, S, or T grid. The density of the R-type grid points can vary by sampling using various random distributions. Additionally, if there is a significant variance in the likelihood of attainability for different input/state pairs, it is recommended to use the trajectory-based S or T grids. Depending on the nature of the system dynamics, the S grid is a proper choice if the attainable subset of the state-domain from steady-state solutions is rich or large enough to ensure coverage of the whole domain by selecting a sufficiently large number of sampling points over each trajectory. On the other hand, this will not be an issue for the T grid, at the expense of including input/state pairs that are only attainable from an unattainable initial state. In general, if such properties of the system dynamics are not fully known, it is suggested to consider all three grid types and compare the overfitting behavior as done in this paper. 
	\item The Kripfganz MMPS form is a compact and well-formulated way to impose continuity in the hybrid approximation of the nonlinear problem; it provides straightforward and intuitive control over the accuracy of the approximation with respect to the number of introduced binary variables that are assigned to each affine local dynamics appearing in the max operators. The number of affine terms can be increased up until the point where either the maximum number of binary variables or the maximum tolerated approximation error are reached. Both of these stopping criteria can be chosen by the user and based on the application. 
	\item The nonlinear non-convex feasible region can be approximated by a union of ellipsoids or polytopes using region-, as well as boundary-based formulations of the approximation problem. If the application requires to strictly avoid violating the nonlinear constraints by the approximated ones, it is recommended to use the region-based formulation of the approximation problem. However, the boundary-based formulation leaves more room to balance the trade-off between covering the nonlinear region and violating it, and converges to better coverage of the non-convex region. This trade-off can also be managed within the region-based formulation by adjusting the tuning parameter $\gamma_{\rm{c}}$, but its capability in modifying the priority of the costs of inclusion vs.\ violation error with respect to the distance from the boundary is limited.
\end{enumerate}

Using the above guidelines, the hybridization approach can be implemented in different applications such as motion planning, navigation, or real-time control of systems with fast dynamics where it is required to balance the computational speed and accuracy of the MPC problem.

\subsection{Next Steps and Future Work}

In the next part of this paper, we present the hybrid control comparison benchmark using this hybridization framework for balancing the computational efficiency of the MPC optimization problem in vehicle control during emergency evasive maneuvers. 

The next steps of the current research can proceed along (but not limited to) the following lines: investigation of the proposed hybridization framework in applications with higher dimensions e.g., large-scale control problems, extension of the model approximation step by incorporating other hybrid modeling frameworks such as piecewise-quadratic or mixed-logical-dynamical systems as compact models for a good trade-off between constraint satisfaction, computational complexity, and control performance.

\section*{Acknowledgment}
 
This research is funded by the Dutch Science Foundation NWO-TTW within the EVOLVE project (no.\ 18484) and by the European Research Council (ERC) under the European Union’s Horizon 2020 research and innovation programme within the CLariNet project (no.\ 101018826). The authors would also like to thank Dr.\ Barys Shyrokau for fruitful discussions on grid generation ideas. 

\bibliographystyle{IEEEtran}
\bibliography{Citations}

\vspace*{-1.5cm}
\begin{IEEEbiography}[{\includegraphics[width=1in,height=1.25in,clip,keepaspectratio]{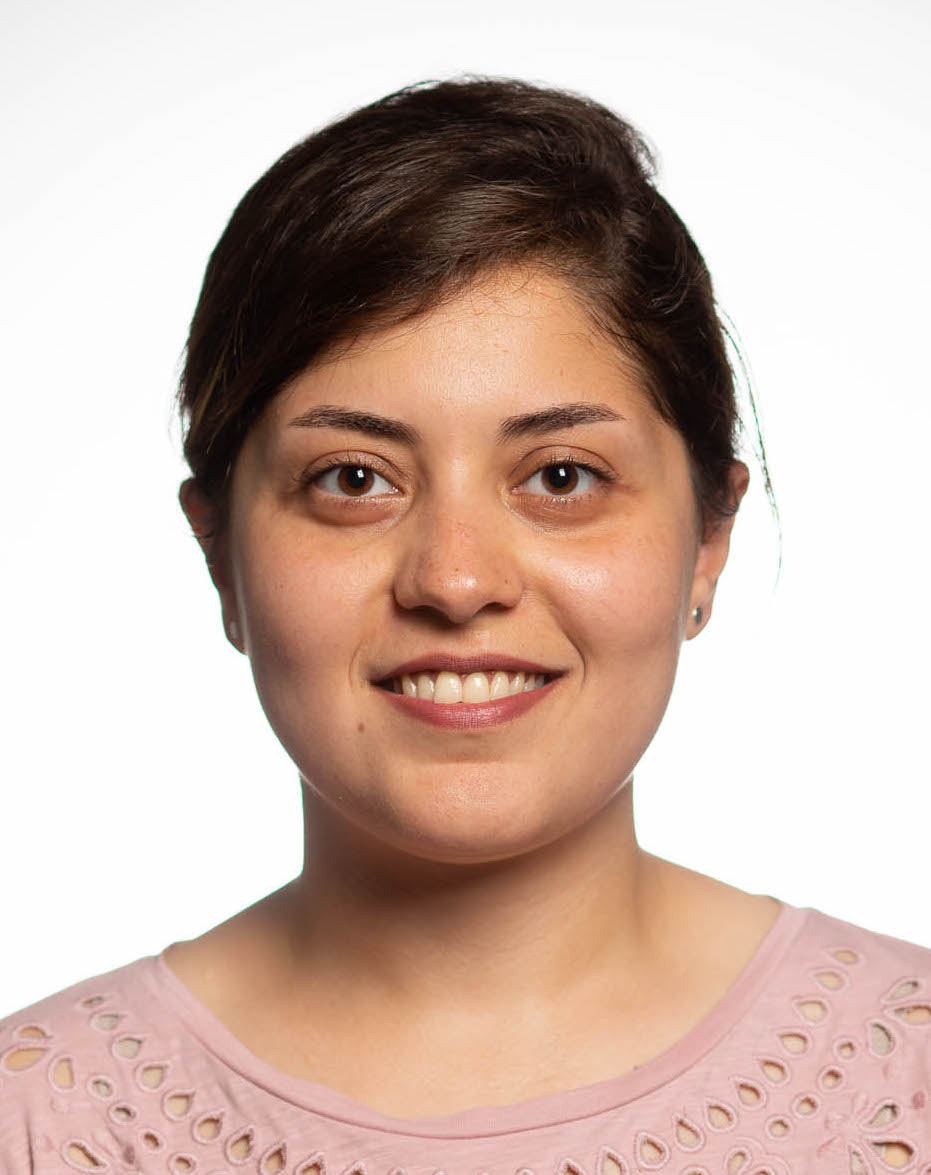}}]{Leila Gharavi} 
	is a PhD candidate at Delft Center for Systems and Control, Delft University of Technology, The Netherlands. She received her BSc and MSc degrees in mechanical engineering from Amirkabir University of Technology (Tehran Polytechnic) in Iran and has research experience in automatic manufacturing and production, vibration analysis and control of nonlinear dynamics, and soft rehabilitation robotics. 
	
	Currently, her research focuses on nonlinear and hybrid systems, optimization, and model-predictive control, with applications to adaptive and proactive control of automated  vehicles in hazardous scenarios.
\end{IEEEbiography}
\vspace*{-1.4cm}
\begin{IEEEbiography}[{\includegraphics[width=1in,height=1.25in,clip,keepaspectratio]{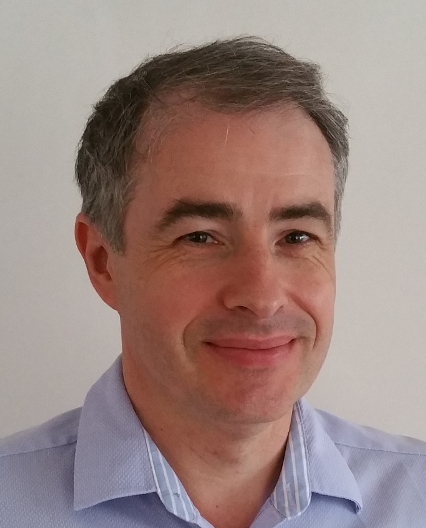}}]{Bart De Schutter}(Fellow, IEEE) 
	received the PhD degree (\emph{summa cum laude}) in applied sciences from KU Leuven, Belgium, in 1996. He is currently a Full Professor and Head of Department at the Delft Center for Systems and Control, Delft
	University of Technology, The Netherlands. His research interests include multi-level
	and multi-agent control, model predictive control, learning-based control, and control
	of hybrid systems, with applications in intelligent transportation systems and smart energy systems. 
	
	Prof.\ De Schutter is a Senior Editor of the IEEE Transactions on Intelligent Transportation Systems and an Associate Editor of the IEEE Transactions
	on Automatic Control.
\end{IEEEbiography}
\vspace*{-1.4cm}
\begin{IEEEbiography}[{\includegraphics[width=1in,height=1.25in,clip,keepaspectratio]{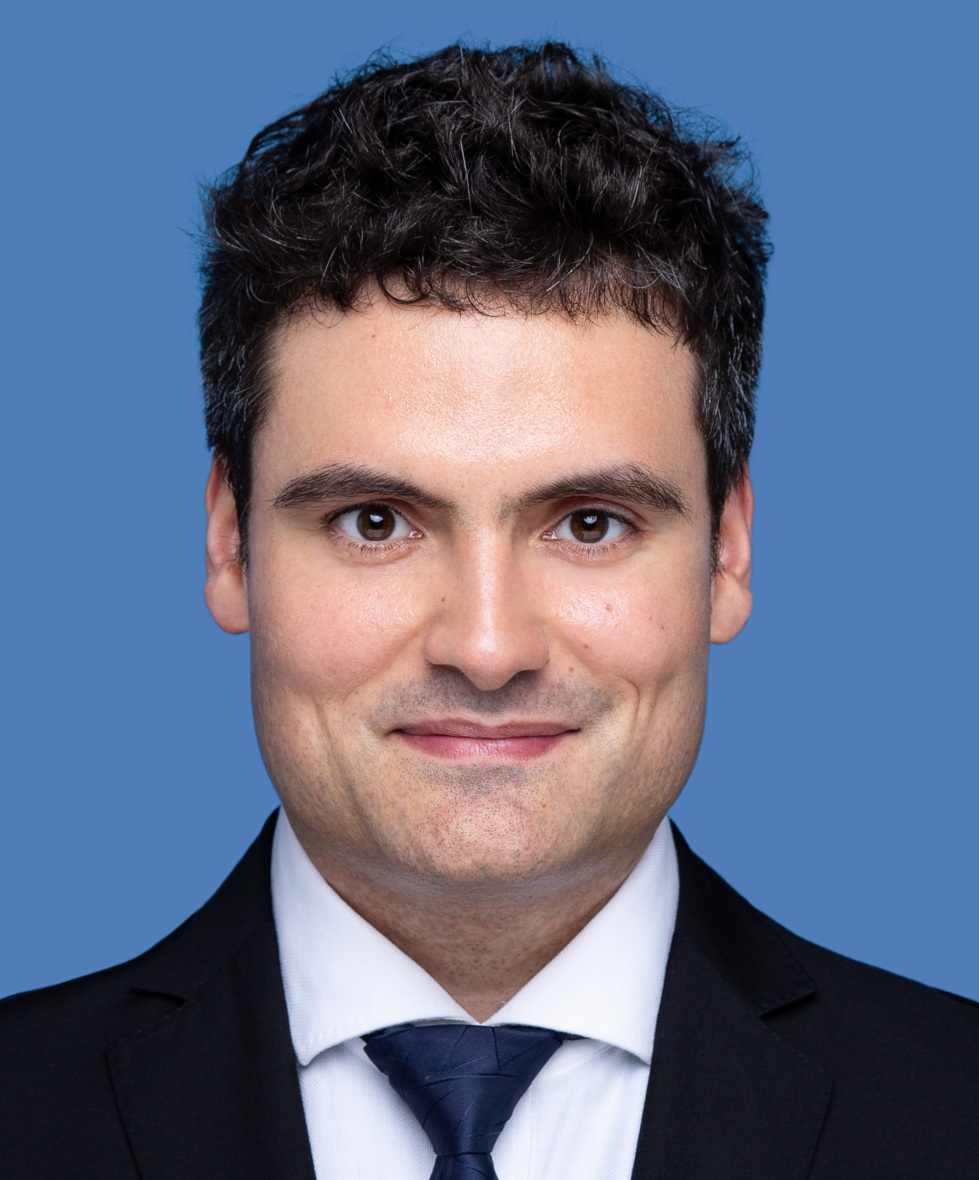}}]{Simone Baldi}
	(Senior Member, IEEE) received the B.Sc. in electrical engineering, and the M.Sc. and Ph.D. in automatic control engineering from University of Florence, Italy, in 2005, 2007, and 2011, respectively. Since 2019, he is a Professor with Southeast University, China, with a guest position with Delft Center for Systems and Control, Delft University of Technology, The Netherlands, where he was Assistant Professor in 2014-2019. His research interests include adaptive and learning systems with applications in intelligent vehicles and smart energy. He was awarded outstanding Reviewer of Applied Energy in 2016, Automatica in 2017, AIAA Journal of Guidance, Control, and Dynamics in 2021. He is a Subject Editor of International Journal of Adaptive Control and Signal Processing, a Technical Editor of IEEE/ASME Transactions on Mechatronics, and an Associate Editor for IEEE Control Systems Letters and Journal of the Franklin Institute.
\end{IEEEbiography}

\end{document}